\documentclass[letterpaper]{article} 
\usepackage{aaai24}  
\usepackage{times}  
\usepackage{helvet}  
\usepackage{courier}  
\usepackage[hyphens]{url}  
\usepackage{graphicx} 
\usepackage{natbib}  
\usepackage{caption} 
\usepackage{algorithm}
\usepackage{algpseudocode}
\usepackage{subcaption}

\usepackage{amssymb}  

\usepackage{amsmath}

\usepackage{todonotes}
\newcounter{todocounter}

\usepackage{newfloat}
\usepackage{listings}
\DeclareCaptionStyle{ruled}{labelfont=normalfont,labelsep=colon,strut=off} 
\floatstyle{ruled}
\newfloat{listing}{tb}{lst}{}
\floatname{listing}{Listing}
\title{vPALs: Towards Verified Performance-aware Learning System \\
  For Resource Management}
\author{
    Guoliang He\textsuperscript{\rm \footnote{Work done as Intern at Huawei Edinburgh Research Centre.},2},
    Gingfung Yeung\textsuperscript{\rm 1},
    Sheriffo Ceesay\textsuperscript{\rm 1},
    Adam Barker\textsuperscript{\rm 1,3},
}
\affiliations {
    \textsuperscript{\rm 1}Huawei Edinburgh Research Centre\\
    \textsuperscript{\rm 2}University of Cambridge\\
    \textsuperscript{\rm 3}University of St. Andrews\\
}

\begin{document}

\maketitle

\begin{abstract}
Accurately predicting task performance at runtime in a cluster is advantageous for a resource management system to determine whether a task should be migrated due to performance degradation caused by interference. This is beneficial for both cluster operators and service owners.
However, deploying performance prediction systems with learning methods requires sophisticated safeguard mechanisms due to the inherent stochastic and black-box natures of these models, such as Deep Neural Networks (DNNs). Vanilla Neural Networks (NNs) can be vulnerable to out-of-distribution data samples that can lead to sub-optimal decisions. To take a step towards a safe learning system in performance prediction, We propose vPALs that leverage well-correlated system metrics, and verification to produce safe performance prediction at runtime, providing an extra layer of safety to integrate learning techniques to cluster resource management systems. Our experiments show that vPALs can outperform vanilla NNs across our benchmark workload.
\end{abstract}

\section{Introduction}

Cluster resource management systems should ensure users' applications run with a satisfactory level of performance. However, the inherent conflicting goal of cloud operators to improve hardware resource utilization, such as sharing the underlying hardware with applications running on the same hardware, ultimately leads to interference between applications, which could lead to significant performance degradation~\cite{zhang2013cpi2,delimitrou2013paragon,llccacheways,wang2021smartharvest}. 

The interference sources can originate from hardware resource contention such as CPU caches, GPU resources, I/0 and network bandwidth~\cite{9428512,romero2018mage,chen2019parties,fried2020caladan}. To anticipate performance degradation at runtime, there has been a large interest in applying machine learning methods to performance prediction~\cite{cilantro,delimitrou2013paragon,grohmann2019monitorless,zhang2021sinan,liu2022nnlqp} for timely performance degradation detection. In particular, Deep Neural Networks (DNNs) can be utilized offline to learn application performance using historic data~\cite{mendoza2021interference,zhang2021nn}. 

To ensure accurate prediction, useful features should be selected to construct the dataset for the training phase. In the context of a cluster resource management system, useful feature sets should contain the system state (usage metrics) collected via observability tools. Hardware usage metrics can provide insights into resource bottlenecks but at the costs of higher overheads~\cite{36575} and delayed aggregation~\cite{kanev2015profiling}. 

Alternatively, system metrics such as CPU and memory usage are proxies of resource bottlenecks and have a lower overhead of collection~\cite{ursa,bashir2021take}. Recent works explore CPU, Memory, and Network usage to predict application performance~\cite{grohmann2019monitorless,zhang2021sinan}. 

Pressure Stall Information (PSI) is a new system-level metric which quantifies in real time how contended a particular CPU, memory or IO resource is. This allows the resource scheduler to maximise the hardware utilization without sacrificing the health of a workload through resource contention. PSI is an interesting candidate metric to predict an application's performance~\cite{weiner2022tmo,lu2023understanding}, however, it remains unexplored to leverage PSI as a feature set to predict an application's performance.

Real computer systems which incorporate DNN predictions can inevitably produce undesirable results due to the probabilistic nature of DNNs~\cite{formal_ver_dnns}, and require system operators to propose safeguard mechanisms in addition to ensuring desirable behavior~\cite{wang2021smartharvest,li2023pond}. It is, therefore, extremely beneficial to deploy a DNN model that is safe and \textit{verified}, in which the DNN performance prediction outputs correctness are guaranteed. 

To address the aforementioned challenges, this paper proposes vPALs, a systematic approach that leverages appropriate system metrics to predict application performance at runtime and combines DNN verification to produce a trustworthy performance prediction model.

This paper makes the following key research contributions:


\begin{itemize}
    \item We have used PSI system-level metrics to learn and accurately predict the performance of various applications. 
    \item We present a novel system that can automate the performance learning and verification process at scale.
    \item We investigate the effectiveness of verified DNNs performance learning in cluster resource management systems.
    
\end{itemize}

\section{Background}
\label{sec:background}

Existing large-scale cluster management systems such as Kubernetes~\cite{burns2016borg} provide mechanisms to collect metrics related to both system and hardware. However, the resource management system does not provide fine-grained hardware metrics information in real-time due to the high overhead of low-level profiling~\cite{36575,wang2022characterizing}. Furthermore, application performance data is not readily available to end users as it is not provided automatically by the cloud provider, and if it is set up within the application, the metrics themselves are often delayed due to the monitoring intervals.

This calls for a lightweight prediction approach that leverages coarse-grained system metrics that correlate well with application performance. The remainder of this section details the background and related work relevant to vPALS.




\subsection{Definitions}

A cluster resource management system e.g., Kubernetes is a software framework that manages a cluster of connected machines $\mathcal{N} = \{n_1,...,n_k\}$. A Job (application) $j$ $\in$ $\mathcal{J}$ comprises one or many \textit{tasks} $\tau_j$. Each task can have one or many replicas $r_{\tau,j}$. A cluster resource management system is responsible for assigning a task replica $r_{\tau,j}$ to a machine $n$. 

During the task replica $r_{\tau,j}$ execution lifetime, the monitoring subsystem collects a set of metrics $M$ from the machine $n$ per timestep $t$, $M_{n,t}$. Our goal is to find a subset $m$ $\subset$ $M$ that can give us a reasonable prediction output that maps $m_{n,t}$ to a task replica performance $p_{r,t}$. Specifically, we aim to learn a function $f$ where $f(m_{n,t}) \approx p_{r,t}$. One of the ongoing challenges of performance prediction is to identify a suitable subset of $m$.

\subsection{Pressure Stall Information (PSI)}

A set of well-correlated metrics $m$ should clearly reflect \textit{resource saturation} in a system at time $t$. The Linux OS kernel exposes a specific set of resource saturation metrics called the \textit{Pressure Stall Information (PSI)}~\cite{psi_linux_kernel}. PSI metrics represent the amount of lost work due to resource pressure or contention in a time interval $t_1 \ldots t_h$, i.e., no resources are available within that time window. These metrics can be measured for a single task replica $r_{\tau}$ or the entire machine $n$. PSI particularly measures the extent to which various resources, such as CPU, memory, and I/O, are being utilized and whether there is resource contention. 

In addition, the PSI metrics have the following sub-measurements, \texttt{some} $PSI_s$ and \texttt{full} $PSI_f$. For example, the \texttt{some CPU} metric $PSI^{cpu}_s$ tracks the percentage of time at least one (runnable) task is stalled due to lack of CPU resources. The \texttt{full memory} $PSI^{mem}_f$ metric tracks the percentage of time when all tasks are stalled due to a lack of memory resources. Note that, CPU PSI metric does not contain \texttt{full} measurements at the system level but exists at the cgroup level because it is impractical that at the system level, all tasks are blocked simultaneously due to lack of CPU resources. Numerous studies show $PSI$ metrics can correctly reflect system pressures in real systems and thus be leveraged as an important metric for various resource management techniques~\cite{weiner2022tmo,lu2023understanding, bannon2018dynamic}. We, therefore, investigate PSI metrics correlation to application performance in the motivation section\ref{sec:motivation}.

\begin{figure}[t!]
    \centering
    \includegraphics[width=\linewidth]{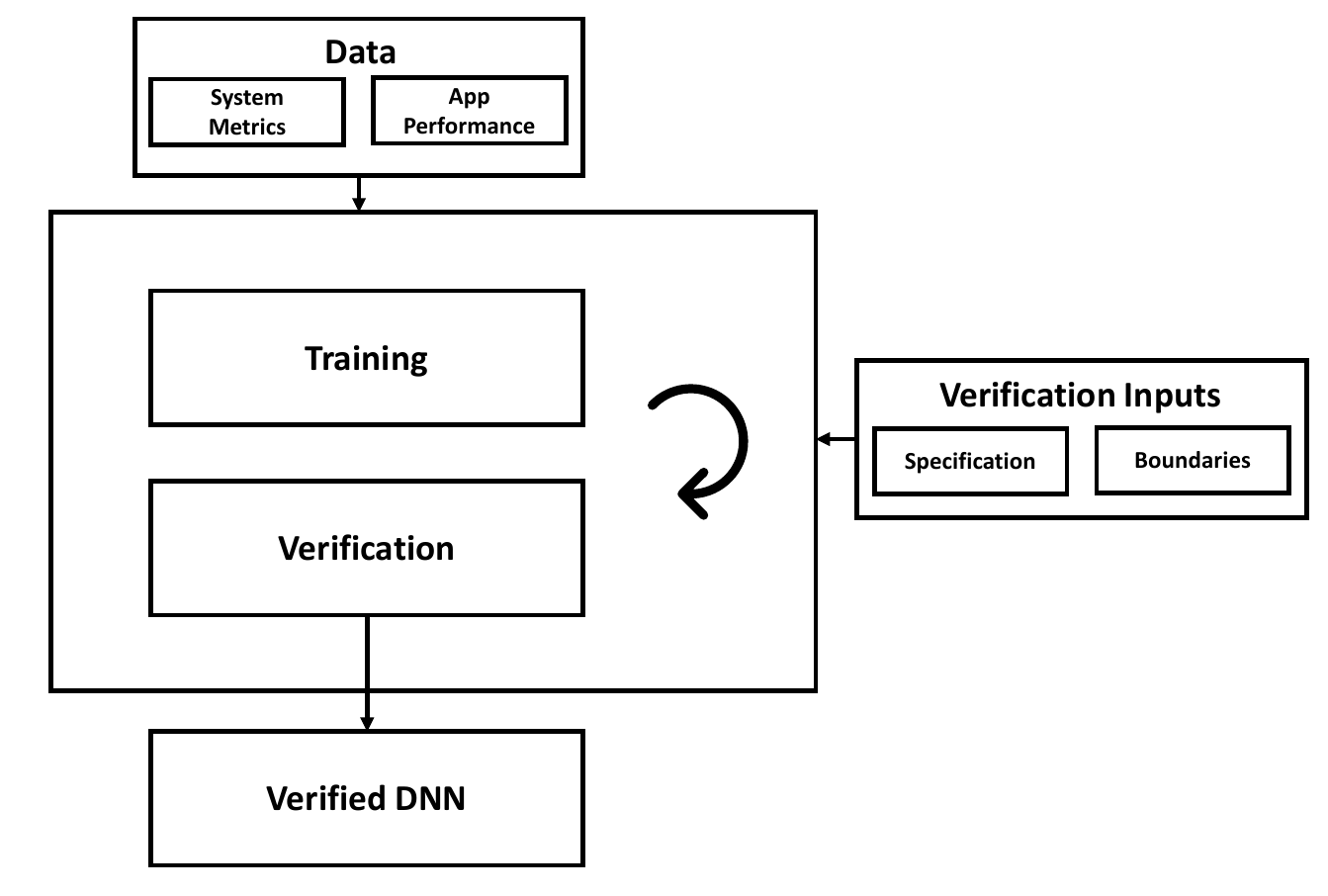}
    \caption{Neural Network Training and Verification System Overview. Performance data and verification bounds are required to produce a satisfactory DNN. }
    \label{fig:perf_train_verify}
\end{figure}

\subsection{Neural Networks Formal Verification}

Deep neural networks (DNNs) are powerful function approximators~\cite{bishop2006pattern,dnn_approx}, and therefore they are widely applied to prediction tasks such as classification~\cite{he2016deep}, language modelling~\cite{touvron2023llama} and regression~\cite{lam2022graphcast}. However, the DNNs performance is usually evaluated via its statistical properties. For example, the percentage of classification accuracy or the mean-square-loss (MSE) of regression tasks is commonly used to train and evaluate the prediction performance. While these criteria are statistically appropriate, it may be frowned upon when system developers or operators try to integrate the DNNs into existing computer systems, where stability and availability are highly expected. For example, 95\% classification accuracy in assigning CPU cores to Virtual Machines indicates 5\% of the time, CPU resource bottlenecks can manifest at run time, resulting in performance degradation. As such, significant system components rely on human heuristics~\cite{wang2021smartharvest,li2023pond} and the prediction of DNNs can only serve as references with confidence upper and lower bounds to guarantee certain performance~\cite{cilantro}. 

Fortunately, recent advances in formal analysis suggest an alternative way to validate DNNs for safe deployment - referred to as \textit{formal verification} \cite{formal_ver_dnns,liu2021algorithms}. A verifier takes as input the verification boundaries, the verification specification, and a DNN to analyze whether the candidate DNN violates the specification in an iterative process as shown in Figure~\ref{fig:perf_train_verify}. In the case of violation, counterexamples are produced and given to the DNN for re-training. Otherwise, the candidate DNN is guaranteed to not violate the specification. With such a verifier, a trained DNN can be safely deployed with trust on its performance, in scenarios such as robotics and computer systems.

Specifically, formal verification for neural networks consists of many specifications, and the three common specifications are i) reachability - our common interpretation of how a neural network should behave, i.e., given an input satisfies a condition, then the corresponding output should satisfy the condition; ii) probabilistic - where a neural network should in addition to obey rules with some given probability, and iii) monotonicity - where the neural network predictions should be able to map the relationships between inputs, this is explained below. In the following definition, the neural network is denoted as a function $f$, its inputs are $x \in \mathcal{D}_x$ and outputs are $y \in \mathcal{D}_y$. For a pair of given input bounds, $\mathcal{X} \subset \mathcal{D}_x$ and $\mathcal{Y} \subset \mathcal{D}_y$, we have

\begin{equation}
    \forall{x_0, x_1}\in \mathcal{X}, \forall{i}, x_0[i]\geq x_1[i] \rightarrow f(x_0)[j] \geq f(x_1)[j]
    \label{eq. monotonic}
\end{equation}

In this paper, we mainly focus on the monotonicity specification, as it is the most relevant to our use case - our verification process is based on \textit{Ouroboros}~\cite{vDNNs} a system designed to train verified DNNs.

\section{Performance Correlation Analysis}
\label{sec:motivation}


Our goal is to find an appropriate subset of metrics $m$ that correlate with a task replica performance $p_{r,t}$. In light of recent publications that reveal the correlation between application performance and PSI metrics, we perform benchmarking experiments to verify the correlation relationship in our compute cluster. Our machines are homogeneous, where each machine runs Ubuntu 20.04 Operating Systems with Linux Kernel 5.4.0-42-generic. Table~\ref{tab:hardware_capacity} describes our machine hardware setup. Our workload consists of \textit{five} common benchmark applications~\cite{cooper2010benchmarking,chen2019parties,sfakianakis2021skynet}, namely \textit{Mindspore}, \textit{MySQL}, \textit{Nginx}, \textit{Redis}, and \textit{Solr}. The workload setup and configurable properties are described in Table~\ref{tab:workload}.

\begin{table}[th!]
\centering
\begin{tabular}{|l|l|}
 \hline
\textbf{Hardware} & \textbf{Infomation} \\ \hline\hline
CPU   & Intel(R) Xeon(R) Gold 6266C CPU \\ 
    &  @ 3.00GHz       \\ \hline
Memory   & DDR4 768GiB  \\ \hline
Disk   & NVMe 3.2 TB        \\ \hline
\end{tabular}
\caption{Machine setup in our compute cluster.}
\label{tab:hardware_capacity}
\end{table}

\begin{table}[t!]
    \centering
    \scriptsize
    \begin{tabular}{|p{1.2cm}|p{1.3cm}|l|l|}
     \hline

        \textbf{Software} & \textbf{Benchmark} & \textbf{Performance} & \textbf{Configurable} \\
        & & \textbf{Metric (ms)} & \textbf{Properties} \\ \hline\hline
        Nginx & Wrk2 & Request  & CPU cores, Memory \\ 
        & & Latency & Rate, Num. Connection  \\ \hline
        Mindspore & Resnet50 & Step Time & CPU cores \\
        Training & Cifar10  &  & \\ \hline
        MySQL & Sysbench & Query & CPU cores, Memory,  \\
         &  &Latency  &  Num. Table, Table size \\ \hline
        Redis & YCSB & Query & CPU cores, Memory \\ 
        & & Latency &   \\ \hline
        Solr & YCSB & Query & CPU cores, Memory \\
        & & Latency &   \\ \hline
    \end{tabular}
    \caption{Benchmark workload used in our experiments.}
    \label{tab:workload}
\end{table}

We need to choose the right level of PSI metric to verify the relationship between application performance and PSI metrics. PSI metrics are measured across CPU, Memory, IO dimensions but can be on a single process level or at the machine $n$. Since we do not want to incur high overhead at runtime during performance prediction, we therefore consider the metrics at the machine $n$ level, $PSI^{res}_{n}$. In addition to the PSI metrics, we also collect the traditional system metrics such as CPU, Network and IO usage for performance learning similar to previous works. Our benchmarking experiment methodology is explained in the Methodology section. In total, we have generated 293 unique colocation combinations of workload for benchmarking. Here we detail our observation on metrics correlation across our applications. 

\begin{figure*}[!t]
   \begin{subfigure}[t]{0.19\textwidth}
     \centering
     \includegraphics[width=\textwidth]{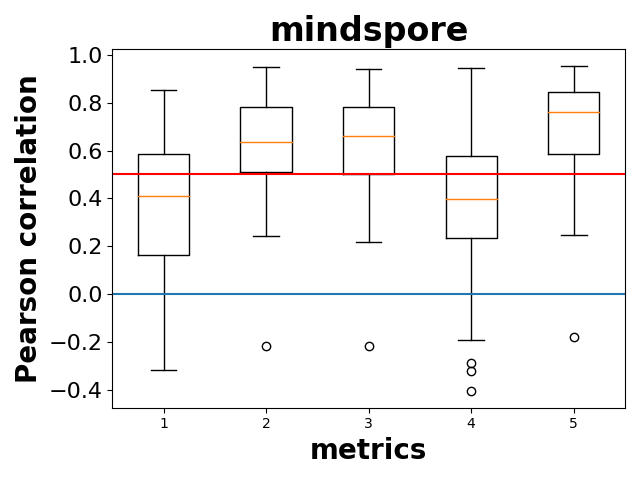}
     \label{fig:mscor}
     \caption{Mindspore correlation.}
   \end{subfigure}
   \hfill
   \begin{subfigure}[t]{0.19\textwidth}
     \centering
     \includegraphics[width=\textwidth]{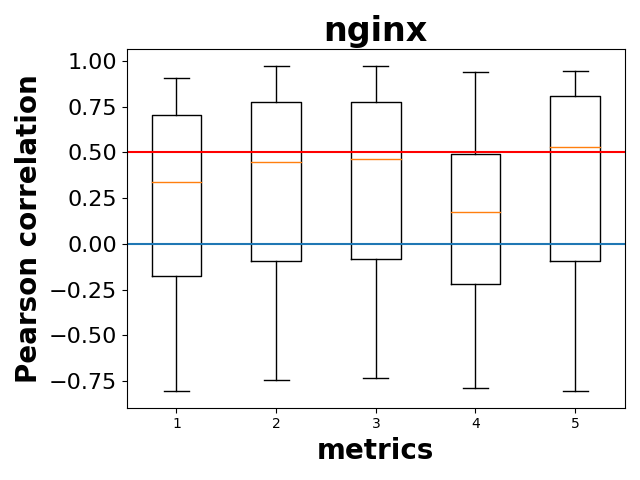}
     \label{fig:nginxcor}
      \caption{Nginx correlation.}
   \end{subfigure}
   \hfill
   \begin{subfigure}[t]{0.19\textwidth}
     \centering
     \includegraphics[width=\textwidth]{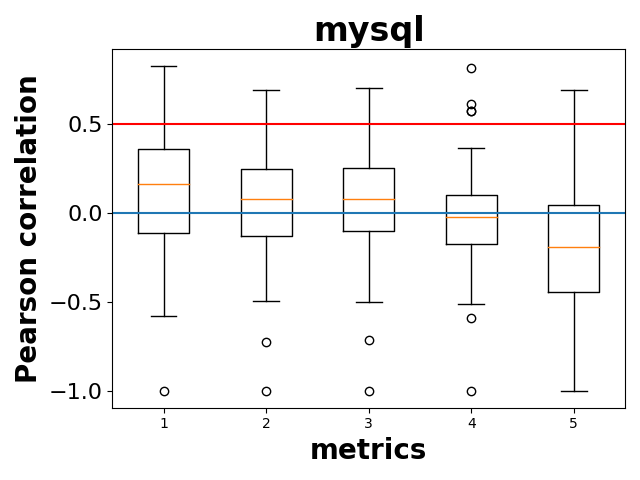}
      \label{fig:mysqlcor}
      \caption{MySQL correlation.}
   \end{subfigure}
   \hfill
   \begin{subfigure}[t]{0.19\textwidth}
     \centering
     \includegraphics[width=\textwidth]{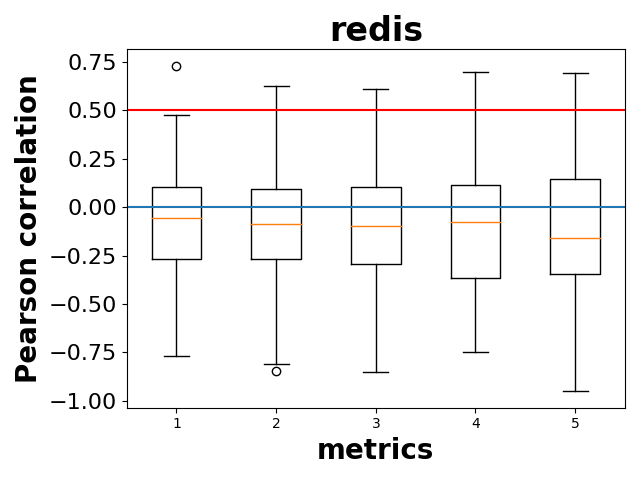}
     \label{fig:rediscor}
       \caption{Redis correlation.}
   \end{subfigure}
   \hfill
   \begin{subfigure}[t]{0.19\textwidth}
    \centering
    \includegraphics[width=\textwidth]{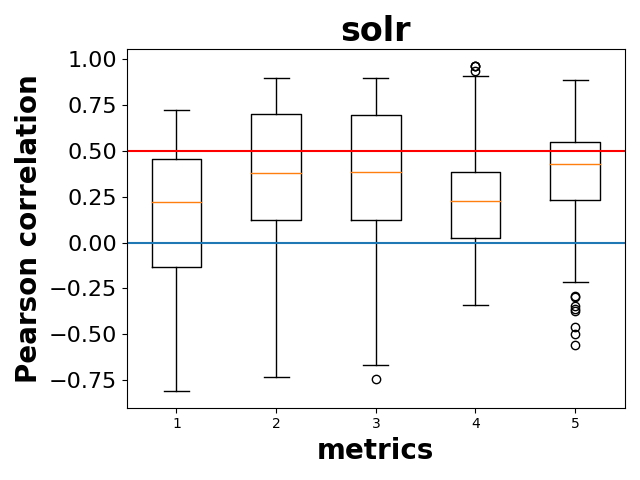}
    \label{fig:solrcor}
    \caption{Solr correlation.}
    \end{subfigure}
   \caption{\textit{Mindspore} and \textit{Nginx} demonstrate a strong correlation between application performance and the system metrics. \textit{Solr} demonstrates a medium correlation, while \textit{Redis} and \textit{Mysql} show a weak correlation between application performance and system metrics. The red line indicates where the Pearson coefficient is $0.5$ and the blue line is $0$.}
   \label{fig:allcor}
\end{figure*}

Across our benchmark, we observe that the most important metrics are CPU, IO, and Disk usage as listed in Table~\ref{tab:metrics_table} and Figure~\ref{fig:allcor}. We omit the other metrics, such as $PSI^{Mem}$ and Network usages, as the variation is too small to be useful. We quantify these remaining metrics relationships via Pearson correlation~\cite{benesty2009pearson} as it is the most common metric to quantify any linear relationships between variables.  Intuitively, by using this metric, we can quantitatively investigate if there exists a positive relationship, i.e., the higher the latency, the higher the system metrics value.

\begin{table}[h!]
    \centering
    \begin{tabular}{|l | l |} 
     \hline
      \textbf{System metrics} & \textbf{ID}    \\ [0.5ex] 
     \hline\hline
     \textit{CPU waiting} ($PSI^{CPU}_s$) & 1  \\ 
     \hline
     \textit{IO stalled} ($PSI^{IO}_f$) & 2  \\ 
     \hline
     \textit{IO waiting} ($PSI^{IO}_s$) & 3  \\ 
     \hline
     \textit{Disk IO time} ($disk$) & 4  \\ 
     \hline
     \textit{CPU time} ($cpu$) & 5  \\ 
     \hline
     \end{tabular}
     \caption{The most significant metrics quantified by Pearson correlation.}
     \label{tab:metrics_table}
\end{table}

Figure \ref{fig:allcor} plots the correlation between the workload application performance and PSI system metrics, respectively, averaged over the runs in our compute cluster. We can observe that $two$ out of $five$ (\textit{Mindspore}, \textit{Nginx}) workloads demonstrate strong positive performance-to-system correlation (correlation coefficient $> 0.5$). While \textit{Mysql}, \textit{Redis} show weak performance-to-system correlation and \textit{Solr} shows a medium correlation relationship. We observe that multiple features (CPU time $cpu$ and IO PSI $PSI^{IO}$) of \textit{Mindspore} and \textit{Nginx} have $ >$ or $ \sim 0.5$ correlation coefficients. The observation shows that application performance can correlate with multiple system metrics instead of only a single type, and a weighting mechanism could be developed to take all system metrics into account for performance estimation. Similarly, for applications like \textit{Mysql} and \textit{Redis}, an individual type of system metrics is a weak indicator to serve as a proxy for the performance metric. The observation implies that a single type of metric to serve as a proxy for the application performance metrics is simply not enough. The performance predictor should consider a significant subset of system metrics $m$ at run time. We note that there are multiple correlation methods to quantify relationships between variables, however, our goal is to demonstrate that PSI + system metrics can be useful for performance learning, hence we leave the discussion and exploration of various correlation methods to future work.

The observation inspires performance learning with multiple types of system metrics. Given that deep neural networks are powerful black-box predictors, we therefore want to learn the application performance end-to-end without developing heuristics to weigh those system metrics. 

At the same time, unlike existing works whose performance learners output prediction with upper and lower bounds \cite{cilantro}, we choose to formally verify our performance learners. This is because verification specification allows us to align our expectation with DNNs prediction, safely deployed into the system, and therefore we do not need to develop extra heuristics to deal with prediction upper and lower confidence bounds.


\section{vPAL System Design}
\label{sec:methodology}

In this section, we will outline the design and implementation of vPALs. We present our data processing pipeline, and how we design our verification specification for the DNNs. 

\begin{figure}[h!]
    \centering
    \includegraphics[width=0.8\linewidth]{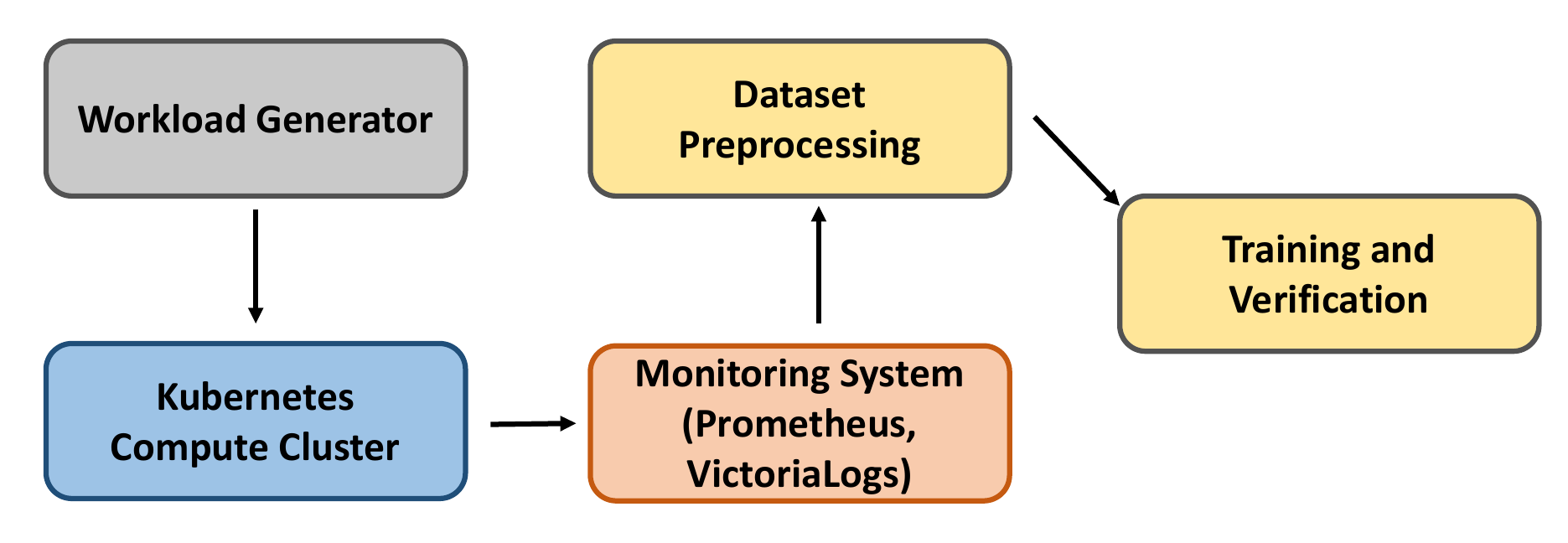}
    \caption{vPALs data pipeline.}
    \label{fig: vpals workflow}
\end{figure}

\subsection{Data processing pipeline}

To generate offline data for training the performance learners, we implement a data processing pipeline to perform Extract-Transform-Load data preprocessing. Note that this is the same processing pipeline for our performance-metrics correlation presented above. The data processing pipeline is general and can be extended to support arbitrary workloads in the future. Currently, we integrate five common workloads as described in Table~\ref{tab:workload}. To manage workload and monitor the applications, we leverage existing open-sourced software for the experiment compute cluster, our software system setup is detailed in Table~\ref{tab:softwaresetup}.

\begin{table}[h!]
    \centering
    \footnotesize
    \begin{tabular}{|l|l|l|}
    \hline
       \textbf{Software}  & \textbf{Description} & \textbf{Version} \\ \hline\hline
       Kubernetes & Resource Management & 1.26.2 \\ 
       \cite{burns2016borg} & Framework &  \\ \hline
       cAdvisor & Monitoring Agent & 0.47.2 \\ 
       \cite{cadvisor} &  &  \\ \hline
       NodeExporter & Monitoring Agent & 1.5.0 \\ 
       \cite{node_exporter} &  &  \\ \hline
       Prometheus & Monitoring Server & 2.42.0 \\
       \cite{prometheus} &  &  \\ \hline
       PromTails & Logging Agent & 2.8.3 \\ 
       \cite{promtail} &  &  \\ \hline
       VictoriaLogs & Logs Storage & 0.3.0 \\ 
       \cite{victoria_logs} &  &  \\ \hline
    \end{tabular}
    \caption{Our compute cluster software and framework setup.}
    \label{tab:softwaresetup}
\end{table}

\paragraph{Experiment Pipeline.} Figure \ref{fig: vpals workflow} depicts our data processing pipeline workflow. We implement a workload generator in Python that interacts with the Kubernetes API server to continuously submit workload to the cluster under certain policies. An example policy includes a modified Monte Carlo simulation~\cite{6968735}, which submits tasks to the cluster as much as possible until the cluster no longer fits or SLA violation. An alternate policy is to co-locate multiple workloads up to a threshold $\theta$ on a compute node to investigate the interference and performance degradation. We use the latter policy to generate benchmarking data, as co-locating workloads up to a certain number is a common approach in the production environment and will create pressure on the compute node. For our experiments, we set $\theta$ to five, as it is a common colocation factor for interference study~\cite{xu2018pythia}.

Our metrics monitoring system includes application logs collection and is deployed to continuously record both application performance and machine system metrics during the experiment process with a configured collection interval. DNN training requires a well-prepared dataset. We discuss our preparation for the dataset below. Specifically, we implement a data cleaning and transformation pipeline to extract data from the metrics logging system. 

\paragraph{Data Cleaning.} Compute cluster workloads typically have variable lifetime~\cite{barbalho2023virtual}. Reasons such as workloads may not get scheduled and assigned as soon as they are submitted to the compute cluster scheduler due to lack of resources or constraints~\cite{zaharia2010delay};  containerized application cold start time which is the initial time to boot up could vary based on the application footprint~\cite{273798}. In our experiments, we try to align the benchmark workload lifetime and limit the cold start time by pre-downloading the images to our benchmark machines. To address the challenges and ensure the data quality, we only keep the data points from the intervals when all workloads are present. We also discard the data points at the beginning of the experiment to account for warm-up time.


\paragraph{Data Transformation.} The data logging intervals have different granularity and timestamps. As a result, we choose to average data points at per-minute intervals, because too fine-grained intervals will introduce system noise, while too coarse-grained intervals may reduce the available data points. In total, the cleansed dataset contains $293$ various application configurations, and $26157$ unique training samples. We perform \textit{MinMax} scaling to pre-process data points and leverage the standard train (80\%) test (20\%) split with random shuffling to prepare the dataset. Formally, each sample data point $x_i$ consists of the subset of metrics $m$ we identified in Table~\ref{tab:metrics_table}, and the target $y_i$ is the target performance. 


\begin{algorithm}[h!]
\caption{Training verified DNNs}\label{alg: vDNNs}
\begin{algorithmic}
\Require Initial dataset $\mathcal{D}$
\Require Initial DNNs
\State Successful $\leftarrow$ false
\While{iteration $< \epsilon$}

\State loss $\leftarrow$ train(DNNs, $\mathcal{D}$)  

\If{loss $<$ theshold \& verifier(DNNs)}
    \State Successful $\leftarrow$ True
    
    \State break

\Else{}
    \State $\mathcal{D} \leftarrow$ $\mathcal{D} +$ counterexamples
\EndIf

\EndWhile
\end{algorithmic}
\end{algorithm}

\subsection{Verification Specification}
\label{sec. verification specification}

We aim to conduct DNN verification before deployment to introduce an extra layer of safety. We leverage an existing DNN verifier, \textit{Ouroboros}~\cite{tan2023building} to train and formally verify our performance learners. We introduce a learner per application as it is a common practice in computing clusters for performance learning~\cite{cilantro}. We leave the unified prediction approach to future work, as our primary goal is to study the DNN verification effectiveness.

An important aspect of DNN verification is the specification. Specifically, we expect the CPU PSI and IO PSI should monotonically decrease with low application latency. Intuitively, \textit{in a computer system, the execution of an application should have lower latency when there are more available system resources and the environment remains unchanged}. Therefore, if the PSI system metrics value is low, we expect the application to have lower latency. 

The monotonicity specification essentially aligns the inference outputs of DNNs according to expert knowledge. The monotonicity verification is defined previously by Equation~\ref{eq. monotonic}. As another example, if two SQL queries process the same database, and one queries a subset of the other, then it is always true that its latency should be smaller given the same environment. In our case, we align the performance learners with our reasoning of application execution and system resources in computing clusters. Based on the correlation and domain knowledge, we design the verification specs tailored to each workload. For example, for \textit{Solr}, its latency is expected to monotonically increase with respect to the second input($PSI^{IO}_f$) and the fourth input($disk$). The rest of the bounds are presented in Table \ref{tab: verification specs detail}. 

\begin{table}[h!]
    \centering
    \footnotesize
    \begin{tabular}{|l|l|l|}
    \hline
       \textbf{Software}  & \textbf{Monotonic features} & \textbf{Increase} \\ \hline\hline
       Mindspore &$PSI^{CPU}_s$, $PSI^{IO}_f$, $PSI^{IO}_s$    & \checkmark \\  \hline
       MySQL &  $PSI^{CPU}_s$, $PSI^{IO}_f$, $PSI^{IO}_s$  & \checkmark \\ \hline
       Solr & $PSI^{IO}_f$, $cpu$  & \checkmark \\  \hline
    \end{tabular}
    \caption{The detailed verification specification for each workload. Redis and Nginx are not considered for reasons listed in the Evaluation section.}
    \label{tab: verification specs detail}
\end{table}

We encode Table \ref{tab: verification specs detail} in \textit{Ouroboros} and leverage its specs-aware mechanism to train verified performance learners. Recall that the verification process iteratively adds counterexamples if the verification fails for a candidate DNN. The addition of counterexamples to the training dataset is the same as in \textit{Ouroboros}. Having defined the verification specification, we can train and produce verified DNNs according to the procedures listed in Algorithm \ref{alg: vDNNs}. 


\begin{figure}[!h]
    \centering
    \includegraphics[width=0.7\linewidth]{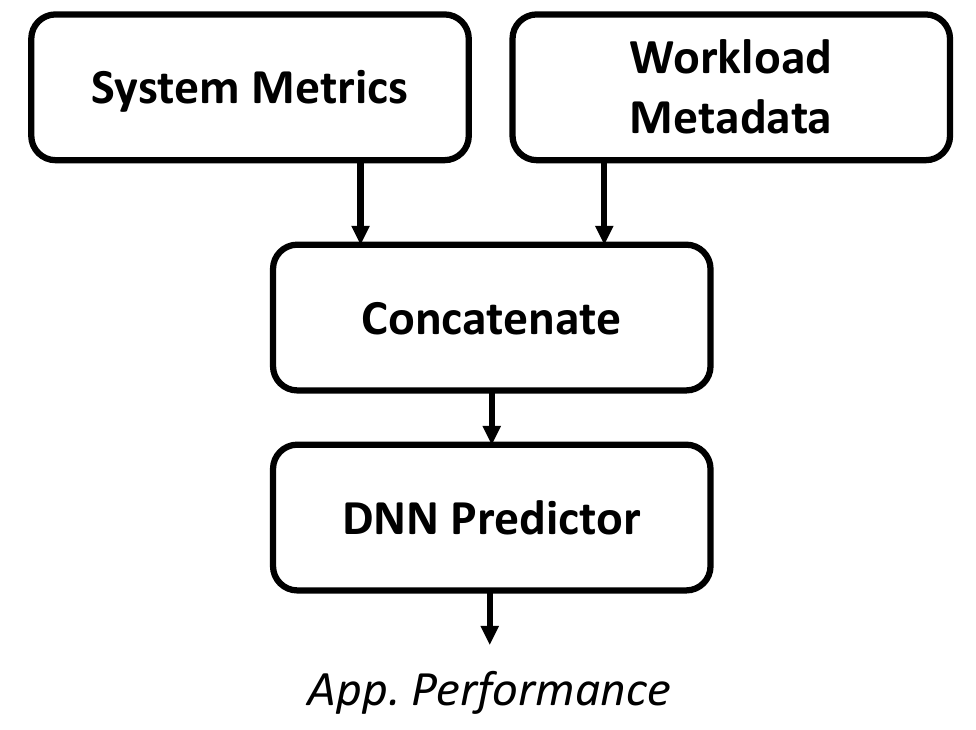}
    
    \caption{Training Overview. The input feature vector consists of a concatenation of system metrics and workload configurations, such as CPU cores and memory requested. }
    \label{fig: training performance learners}
\end{figure}

\begin{figure*}[t!]
\begin{subfigure}[t]{0.32\textwidth}
     \centering
     \includegraphics[width=\textwidth]{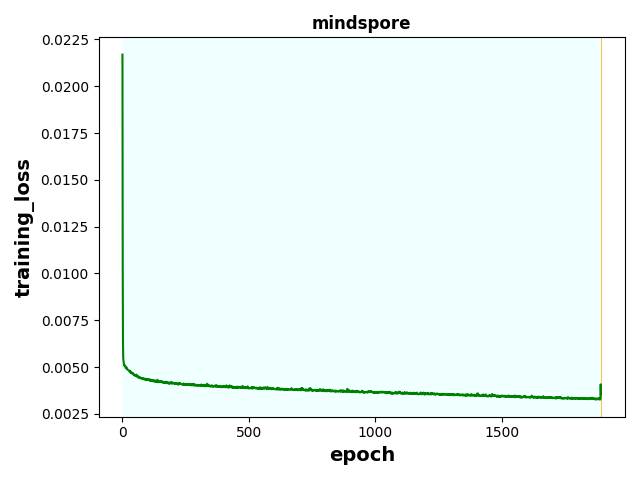}
     \label{fig:msloss}
     \caption{Mindspore Trainning Loss.}
   \end{subfigure}
    \hfill
\begin{subfigure}[t]{0.32\textwidth}
     \centering
     \includegraphics[width=\textwidth]{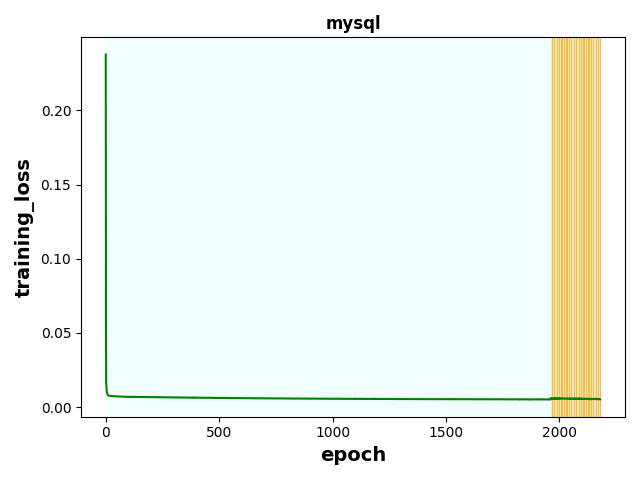}
     \label{fig:msqlloss}
     \caption{MySQL Training Loss.}
   \end{subfigure}
    \hfill
\begin{subfigure}[t]{0.32\textwidth}
     \centering
     \includegraphics[width=\textwidth]{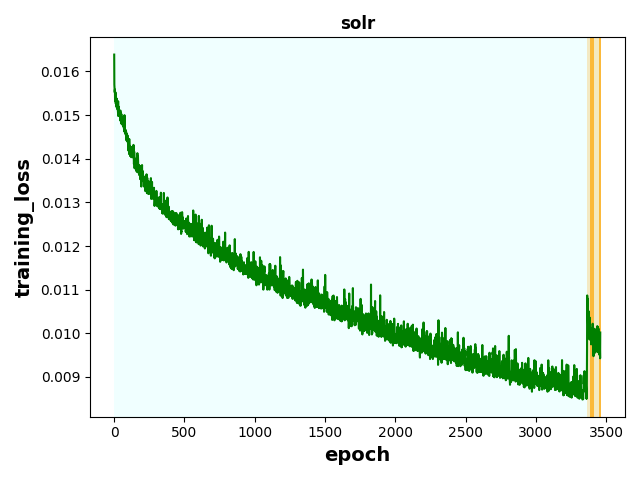}
     \label{fig:solrloss}
     \caption{Solr Training Loss.}
   \end{subfigure}
\caption{Training loss versus epochs. The yellow interval is when the verifier activates and performs verification.}
\label{fig. training loss versus epochs}
\end{figure*}

\paragraph{DNN Training.} Our candidate DNNs are trained using both system metrics and workload metadata. The input features are concatenated as a feature vector. We follow the practice of incorporating domain-specific knowledge, as it has been shown that leveraging domain-specific features can outperform generic feature set~\cite{liu2022nnlqp,mendoza2021interference}. Furthermore, the application-specific DNNs are designed to be efficient and simple, this is due to the design consideration that for any performance learners that should run in real-time should be as lightweight as possible and not incur high overhead to the system, i.e., creating more resource contention. The application-specific DNNs only differ across the input dimensions where the metadata features are different. Specifically, our application-specific DNN model's intermediate layers and hidden dimensions are described in Table~\ref{tab:dnns_definition}. We leverage ReLU~\cite{krizhevsky2012imagenet} activation function and Adam~\cite{kingma2014adam} Optimizer with a learning rate of $1e^{-4}$ for our DNNs.

\begin{table}[h!]
    \centering
    \small
    \begin{tabular}{|l|l|l|}
    \hline
         \# Layers & Hidden Dimensions & Workloads   \\ \hline\hline
         2 & 300 & Mindspore \\
         &   &  MySQL, Redis \\ \hline
         3 & 300 & Nginx \\ \hline
         3 & 512 & Solr \\ \hline
    \end{tabular}
    \caption{Fully Connected DNNs Definition.}
    \label{tab:dnns_definition}
\end{table}


\begin{figure*}[t!]
\begin{subfigure}[t]{0.32\textwidth}
     \centering
     \includegraphics[width=\textwidth]{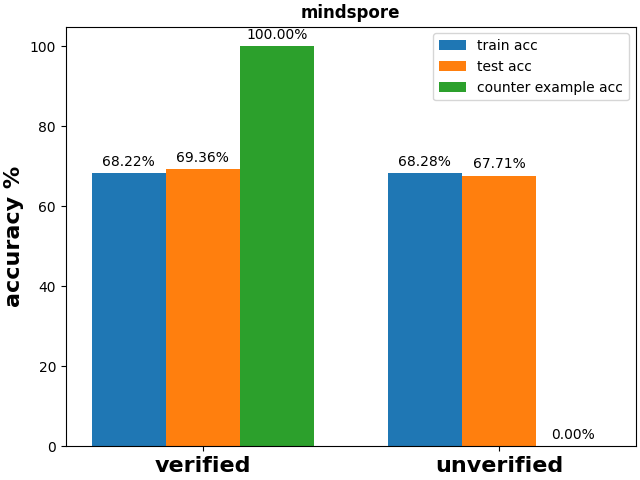}
     \label{fig:ms-acc}
   \end{subfigure}
    \hfill
\begin{subfigure}[t]{0.32\textwidth}
     \centering
     \includegraphics[width=\textwidth]{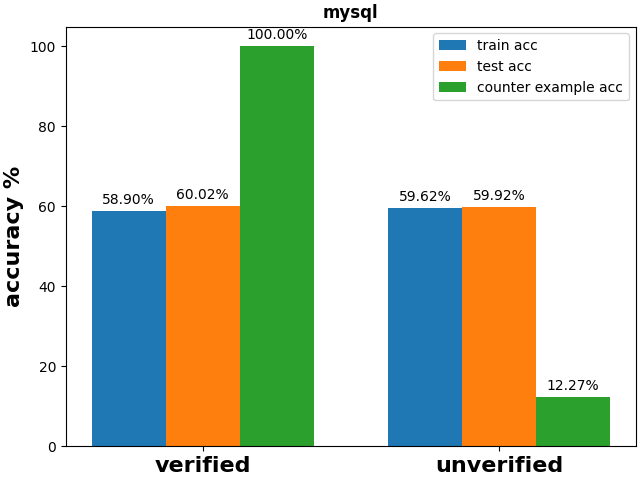}
     \label{fig:msql-acc}
   \end{subfigure}
    \hfill
\begin{subfigure}[t]{0.32\textwidth}
     \centering
     \includegraphics[width=\textwidth]{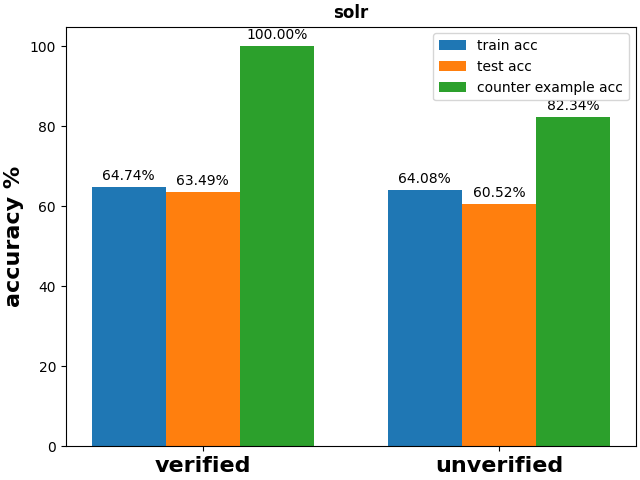}
     \label{fig:solr-acc}
   \end{subfigure}
   
\caption{Accuracy comparison between verified and unverified DNNs. Verified DNNs outperform unverified DNNs by a margin for counterexamples set and $\sim3\%$ on the test set.}
\label{fig. ver versus unver}
\end{figure*}

\section{Evaluation}
\label{sec. Evaluation}

We evaluate vPALs with the specified specification bounds (Table~\ref{tab: verification specs detail}),  to answer the following questions:

\begin{itemize}
    \item How does verification help performance learners?
    \item What are the limitations of performance prediction using verification?
\end{itemize}

\subsection{Experimental Setup}
\label{sec. Experiment setups}

\paragraph{Comparison Baselines.} We follow existing verification literature where we compare our verified DNNs with DNNs without verification to demonstrate potential benefits. Specifically, after we obtain a verified DNN for a specific application at a particular epoch $e$, we also train a corresponding candidate DNN without verification up until the specific epoch regardless of the training loss. Note that, we set the same random seed for both training processes. Finally, we also set a $e_{lim}$ for the verification process to ensure only reasonable efforts are spent on it. The limit is set to 5000. Thus, we can faithfully compare the two candidate DNNs with the same training process but only differ in whether they participate in verification.

\paragraph{Metrics.} We use Mean-Square-Error (MSE) as the loss function to train our DNNs, as our performance learner predicts application latency directly, i.e., a regression task. However, MSE is not very intuitive when it comes to reasoning about the performance of prediction. As a result, we report an additional accuracy $acc$ metric to quantify the result, this is shown as the following:

\begin{equation}
    acc := abs(f(x) - y) < 0.05
\end{equation}

Therefore, the overall accuracy metric is simply the percentage of correct prediction in the data set. The accuracy threshold ($0.05$) is a hyperparameter, and we use the same value as leveraged in \textit{Ouroboros} for regression tasks.


\subsection{Training process and verification}

Figure \ref{fig. training loss versus epochs} shows the training loss, and the verification stages, denoted by the yellow interval. Note that, in our benchmarking, the performance of \textit{Nginx} is stable across our data samples with practically no differences in latencies, hence we omit the DNN performance learning. For \textit{Redis}, the training loss cannot converge, and therefore we also omit the figure. A potential reason could be due to the metrics we selected are of low correlation, which is observed in Figure~\ref{fig:allcor}.

We can observe that as soon as the verifier is activated, the training loss spikes. This is because counterexamples are added to the training set, and as the name suggested, the performance learner mistakenly predicts the latency of counterexamples, resulting in the increment of training loss. As more training iterations proceed, the training loss gradually decreases again, indicating the performance learner learns to predict the latency correctly even with counterexamples. Note that the verifier is activated as soon as the training accuracy is lower than a pre-defined threshold. We set a threshold per application, following the same practice as \textit{Ouroboros}. 

We can observe that when there are stronger correlations with the subset of metrics we identified, it the easier to verify the DNNs. As we demonstrated in Figure \ref{fig:allcor}, \textit{Mindspore} has a strong performance-to-system correlation, and as a result, the verifier successfully verifies within a few iterations. On the other hand, \textit{MySQL} has a weak performance-to-system correlation, and therefore it takes a few hundred verification iterations to verify, while \textit{Solr} needs tens of epochs to be verified. As an extreme example, \textit{Redis} cannot converge within a reasonable number of epochs. We discuss the limitations of our approach in a later section.

This phenomenon may present a new future direction to understanding the verifiability of DNNs. High-quality data with strongly correlated features update the DNNs towards a region where fewer counterexamples are located. As a result, the verifier does not need to add many counterexamples to adjust the dataset and the performance learner can be verified easily. This is intuitive, as a well-balanced and well-correlated dataset that covers the entire distribution will inevitably cover the verification specification. 


\subsection{Verified Neural Networks Performance}

We compare the traditional (unverified) DNNs and verified DNNs performance on the training, test, and counterexample sets, respectively. To ensure fair evaluation, we prepare the datasets similarly and train both networks with the same number of epochs. Figure \ref{fig. ver versus unver} shows the result of the comparison. We observe that the two DNNs types' performances in the training set and test set are similar, with the verified DNNs having slightly higher accuracy in the test set $\sim3\%$, which shows the verification does not degrade the model performance but improves it slightly. 

On the other hand, the verified DNNs achieve $100\%$ accuracy in the counterexamples set, which is reasonable because the train-and-verify loop guarantees that the trained performance learner always satisfies the verification specification. However, for unverified DNNs, the \textit{Mindspore} performance learner violates all counterexamples, while the \textit{MySQL} performance learner only predicts $12.5\%$ counterexamples correctly. Surprisingly, the \textit{Solr} learner achieves $82.3\%$. This indicates if we deploy unverified DNNs to online systems, there will be cases where even though the system resources are limited, the performance learner could predict a low latency, which may result in violation of SLAs or inefficient resource management. 

%

%

%

\section{Limitations}

There are two main limitations of vPALs. 

\begin{itemize}
    \item Existing verifiers cannot scale beyond several layers of MLP. While this is not an issue at the moment, as larger networks may consume more compute resources and bring inference overhead, if we were to leverage state-of-the-art models (e.g., large language models) for performance learning in the future, the verifier will not be able to verify within a finite amount of time.
    \item The performance learners assume to access workload metadata and use it to construct the input vector. While this is achievable at runtime, the workload configuration may be out-of-distribution, and we rely on the generalization ability of DNNs to make predictions. We leave the generalizability investigation to future work. 
\end{itemize}

\section{Future Work and Conclusion}

\paragraph{Unified learner.} Our initial approach requires a learner per application, we plan to investigate a unified approach to enhance generalizability. One potential approach to develop a unified learner by leveraging a shared backbone that takes in intermediate features from domain-specific heads. 



\paragraph{Uncertainty-aware.} Uncertainty-aware learning approaches are beneficial in producing outputs containing both the upper and lower bound values~\cite{gal2016dropout}. A potential direction is to include verification for uncertainty-aware learning methods, i.e., guarantee the produced upper and lower bound values do not fluctuate. We plan to also compare against uncertainty-aware learning approaches or active learning methods to aid the safe deployment of AI in the future.


\paragraph{Conclusion.} We present vPALs, a step towards performance learning for resource management systems. We show that PSI metrics are indeed an important set of features for performance learning. We show that verification does not degrade performance and even slightly improves accuracy. We show that vanilla DNNs without verification will fail in most counterexamples in our dataset, thus verification for performance learning can add another safety layer in deploying AI in resource management systems.

\bibliography{aaai24}

\end{document}